\documentclass{emulateapj}

\shorttitle{IGR J22517+2218=MG3 J225155+2217: a new gamma-ray lighthouse in the distant Universe}
\shortauthors{Bassani et al.}

\begin{document}

\title{IGR J22517+2218=MG3 J225155+2217: a new gamma-ray lighthouse in the distant Universe}

\author{L. Bassani\altaffilmark{1},  R. Landi\altaffilmark{1}, A. Malizia\altaffilmark{1}, M.T. Fiocchi\altaffilmark{2},
A. Bazzano\altaffilmark{2},
A. J. Bird\altaffilmark{3}, A. J. Dean\altaffilmark{3},  N. Gehrels\altaffilmark{4}, \\
P. Giommi\altaffilmark{4}, P. Ubertini\altaffilmark{2}}
\altaffiltext{1}{IASF-Bologna/INAF, Via Gobetti 101, I-40129 Bologna, Italy}
\altaffiltext{2}{IASF-Roma/INAF, Via Fosso del Cavaliere 100, I-00133 Rome, Italy} 
\altaffiltext{3}{School of Physics and Astronomy, University of
Southampton, Highfield, Southampton, SO 17 1BJ, UK.}
\altaffiltext{4}{NASA/Goddard Space Flight Center, Greenbelt, MD 20771}
\altaffiltext{5}{ASI Science Data Center, ESRIN, I-00044 Frascati, Italy}

\begin{abstract}
We report on the identification of a new soft gamma ray source, namely IGR J22517+2218, detected with IBIS/INTEGRAL. The source, which  has an observed 20-100 keV flux of $\sim$4 $\times$ 10$^{-11}$ erg cm$^{-2}$ s$^{-1}$, is spatially coincident with
MG3 J225155+2217, a quasar at  z=3.668. The Swift/XRT 0.5-10 keV continuum is flat ($\Gamma$=1.5) with evidence for a spectral curvature below 1-2 keV either due to intrinsic absorption (N$_{H}$=3$\pm$2 $\times$ 10$^{22}$ cm$^{-2}$) or 
to a change in slope ($\Delta\Gamma$= 0.5). 
X-ray observations indicate flux variability over a 6 days period  which is further supported by a flux  mismatch between Swift and INTEGRAL spectra. IGR J22517+2218 is radio loud and has a flat radio spectrum; 
optically it is a broad line emitting  quasar with the atypical
property of hosting a narrow line absorption system. The Source Spectral Energy Distribution is unusual compared to blazars of similar type: either it has the synchrotron peak in the X/gamma-ray band (i.e. much higher than 
generally observed) or  the Compton peak in the MeV range (i.e. lower than  typically measured).
IGR J22517+2218=MG3 J225155+2217 is the second most distant blazar detected above 20 keV and  a gamma-ray lighthouse shining from the edge of our Universe. 
\end{abstract}


\section{Introduction}
Blazars are the most powerful of all AGN seen in the observable Universe; 
their continuously emitting radiation covers the entire electromagnetic spectrum from radio to gamma-ray frequencies. 
Because of their enormous luminosities, blazars are visible to very large distances/redshifts.
In the widely adopted scenario of blazars, a single population of 
high-energy electrons in a relativistic jet radiate from the radio/FIR to 
the UV- soft X-rays by the synchrotron
process and at higher frequencies by inverse Compton scattering soft-target 
photons present either in the jet, in 
the surrounding material  or in both (Ghisellini et al. 1998). 
Therefore a strong signature of the blazar nature 
of a source is a double peaked structure in the Spectral Energy Distribution 
(SED), with the synchrotron component peaking anywhere from infrared to 
X-rays and the
inverse Compton extending up to GeV/TeV gamma-rays.
To explain all the different SEDs observed in blazars, Fossati et al.
(1998) proposed the blazar's sequence, which claims an inverse relation
between peak energies and source luminosity: the more luminous sources
have both synchrotron and Inverse Compton peaks at lower energies than
their fainter (and generally at lower redshift) counterparts.
Within the blazars population, high redshift objects, which  belong to the
class of flat spectrum radio Quasars (FSRQ), tend to be the most
luminous one. Their SED peak at IR/optical (first peak) and GeV (second
peak) frequencies and only a handful of objects show both peaks at lower
energies (Blom et al.  1996). 
Furthermore, FSRQ with a synchrotron peak above the
optical band are
extremely rare (Giommi et al. 2007) and not predicted by the blazar's
sequence. In less luminous blazars, on the other hand, the first peak can reach the $\sim$100 keV band
while the second one goes well into the TeV range.
Observations above 10 keV are therefore extremely important as
they can select rare type of blazars, i.e. those with the synchrotron or
Inverse Compton  peak
at hard X-ray/ gamma-ray energies. Finding either type of objects is
crucial to test the validity of the blazar's sequence and to study jet
parameters in extreme blazars or, in other words, to stretch blazar
theories to the limit. Unfortunately, observational progress 
in Blazar's studies above 10 keV has been slow, particularly for the most distant sources.
Nonetheless, a few objects at high z have been detected by 
BeppoSAX/PDS  first (less than 10 in the 1-4.7 z range, Donato et al. 2005) and more recently by 
INTEGRAL/IBIS (Bird et al. 2007) and Swift/BAT (Sambruna et al. 2007). 
Here, we provide  evidence for the 
identification of a newly discovered INTEGRAL/IBIS source,
IGR J22517+2218, with a quasar (QSO) at z=3.668 by means of  follow up observations with Swift/XRT; 
we also present the  combined X/gamma-ray spectrum of the source and discuss its peculiar SED.
This is the  second most distant blazar ever detected above 20 keV.

\section{IGR J22517+2218}
IGR J22517+2218 was first reported by Krivonos et al. (2007)
as an unidentified object detected during revolution 316 (150 ksec exposure).
Here, we have combined slightly more data to reach  
a total exposure of 190 ks. IBIS data have been reduced following the same procedure used
for our survey work (Bird et al. 2007). The
source is detected with a significance of $\sim$7$\sigma$ 
at  R.A.(2000)=22h51m42.72s
and Dec(2000)=+22$^{\circ}$17'56.4" with a positional uncertainty
of  4.5' (90\% confidence level see Figure.1); this position is 32$^{\circ}$ above the galactic plane, 
suggestive of an extragalactic nature.
Our position is compatible with that reported by Krivonos et al (2007) (note that their 2.1' error radius corresponds to 68\% confidence level).
The source count rate  in the 20-100 keV band is 0.4 $\pm$0.06 counts s$^{-1}$ (or 2.17$\pm$0.33 mCrab). 
We have  divided all available observations in 3 sets of data 
to find evidence of flux variability  but found none.
The IBIS spectrum was extracted using the standard Off-line Scientific Analysis
(OSA version 5.1) software released by the Integral Scientific Data Centre.
Here and in the following, spectral
analysis was performed with XSPEC v.11.3.2 package and errors are
quoted at 90\% confidence level for one interesting parameter
($\Delta\chi^{2}=2.71$).
A simple power law provides a
good fit to the IBIS data ($\chi^2$=6.5 for 8 d.o.f.)
and a photon index $\Gamma$=1.4$\pm$0.6 combined to an observed  20-100 keV 
flux of 4 $\times$ 10$^{-11}$ erg cm$^{-2}$ s$^{-1}$.
Within the INTEGRAL/IBIS error circle, we find the QSO MG3 J225155+2217 at z=3.668; however its association with IGR J22517+2218  is not straightforward as 1) there are other potential candidates in the 
high energy error box  for example at radio frequencies
and 2) it is surprising to detect such a distant object in a short exposure with INTEGRAL.
For this reason we have requested and obtained  follow up observations of this sky region with the XRT telescope (Hill et al. 2004) on board the Swift satellite (Gehrels et al.2004).

\begin{figure}
{\epsscale{0.85} \plotone{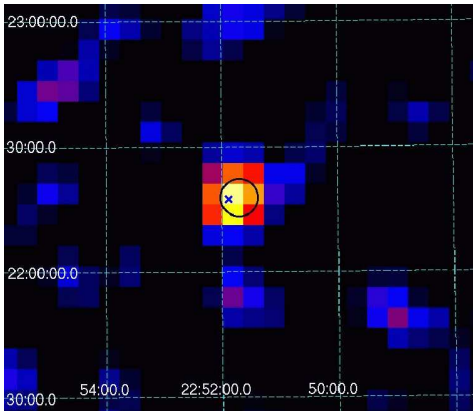} }\caption {IBIS/ISGRI 20-100 keV image of  IGR J22517+2218. 
The circle corresponds to the INTEGRAL/IBIS error box while the X is the optical position of MG3 J225155+2217. \label{fig1}}
\end{figure}

\section{SWIFT/XRT follow up observation}

On  May 21, 22 and 26,  2007  Swift/XRT carried out 
observations of IGR J225155+2217 lasting 1.6, 12.5 and 11.3 ks respectively.
Data were collected in Photon Counting mode and
reduction was performed using the XRTDAS v2.0.1 standard data pipeline package.
Events for spectral analysis were extracted within a circular region of radius 20$^{\prime \prime}$ 
(which encloses about 90\% of the PSF at 1.5 keV, Moretti et al. 2004) 
centered on the source position.
The background was taken from various source-free regions close to the 
X-ray source  using  circular/annular regions with 
different radii, in order to ensure an evenly sampled background. In all 
cases, the spectra were extracted from the corresponding event files using 
{\sc XSELECT} software and binned using {\sc grppha} in an appropriate 
way, so that the $\chi^{2}$ statistic could reliably be used. We used the 
latest version (v.008) of the response matrices and create individual 
ancillary response files (ARF) using {\sc xrtmkarf}. 
In all our fitting procedures we have used a Galactic 
column density  which in the direction of IGR J225155+2217 is 5 $\times$ 10$^{20}$ cm$^{-2}$ (Dickey \& Lockman, 1990).
Using the XIMAGE detection algorithm, we searched the  XRT images (0.3-10 keV band)  for
significant excesses (above 3 sigma level)  falling within the INTEGRAL/IBIS 90\% confidence
circle (figure 2); only one source (N1) was clearly detected in all 3 observations 
(at 10, 27 and 20 $\sigma$) 
while a second one (N2) was seen (at 4.5 $\sigma$) only in the longest XRT exposure. None of the other potential candidates in the IBIS error box is detected in X-rays. Table 1 reports for source N1 and N2 the XRT position,  
associated error radius and, for each observation, net count rates and fluxes. Source N1 is the high redshift QSO MG3 J225155+2217.
Its XRT spectrum is hard and bright: a simple power law fit to the two longest 
observations provides $\Gamma$=1.41$\pm$0.12 ($\chi^{2}$=28.9, 17) and $\Gamma$=1.30$\pm$0.08 ($\chi^{2}$=34.4, 41 dof) together with a  2-10 keV flux in the range 2.6-3.4 $\times$ 10$^{-12}$ erg cm$^{-2}$ s$^{-1}$.
The XRT spectrum of  source N2 is, instead, very soft as no emission is detected above 3 keV; it is well fitted by a 
thermal bremstrahulung model with kT=0.4$^{+0.1}_{-0.3}$ ($\chi^{2}$=4, 3 dof) and a 2-10 keV flux of
$\sim$ 10$^{-15}$ erg cm$^{-2}$ s$^{-1}$.  Its positional coincidence with either a double (CCDM J22519+2219AB) or a single (TYC 1710-969-1) star, suggests  that the X-ray emission comes from a stellar corona;  the late spectral type
(F5) of CCDM J22519+2219AB is fully compatible with this hypothesis. In any case, the much lower flux (more than  3 orders of magnitudes) and softer spectrum of source N2 compared to source N1 indicates that its emission is unlikely to extend into the INTEGRAL energy range and strongly argues in favour of the association of IGR J225155+2217 with the only other detected X-ray counterpart, i.e. the high z QSO.

\begin{figure}
{\epsscale{0.85} \plotone{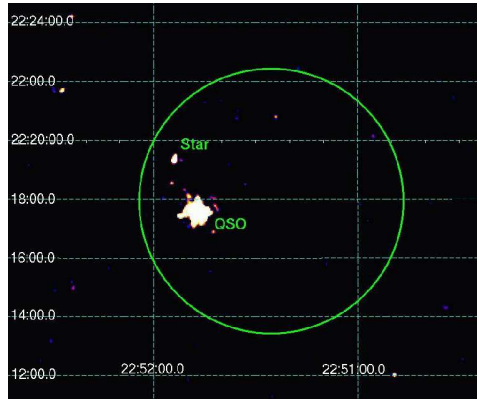}} \caption {Swift/XRT (2-10 keV) image of IGR J225155+2217. The circle 
corresponds to the INTEGRAL/IBIS error box.  \label{fig2}}
\end{figure}

\begin{table*}
\begin{center}
\footnotesize
\caption{\emph{Swift}/XRT detections}
\begin{tabular}{lcccccc}
\tableline
\tableline
 Source         &      RA$^{\dagger}$     &    Dec$^{\dagger}$        & error$^{\dagger}$  & CR/ Flux(obs1)$^{\ddagger}$ & CR/ Flux(obs2)$^{\ddagger}$ & CR/ Flux(obs3)$^{\ddagger}$   \\
 \tableline
N1 & 22 51 53.44  &   +22 17 35.86   & 3.6 &  8.77$\pm$0.85/ 4.4$\pm$0.4  & 8.36$\pm$0.31/ 3.4$\pm$0.1             &   7.05$\pm$0.35/ 2.6$\pm$0.1 \\  
N2 & 22 51 56.70  &  +22 19 22.90    & 4.6 &  0.59$\pm$0.22/ $\le$0.003     & 0.21$\pm$0.06/ 0.001$\pm$0.0002  &  0.18$\pm$0.06/ $\le$ 0.001    \\     
\tableline
\end{tabular}
\end{center}
Notes:$^{\dagger}$ Positions and relative uncertainties (in arcsec) are calculated using the xrtcentroid (v0.2.7) task.
$^{\ddagger}$ Count rates (CR) are extracted from 0.3-10 keV band image using the XIMAGE detection algorithm and are in units of 10$^{-2}$ cts/s in the  band; fluxes are in units of 10$^{-12}$ erg cm$^{-2}$ s$^{-1}$ in the 2-10 keV band. 
\end{table*}

To better characterized the QSO X-ray spectrum  and in view of the fact that the  INTEGRAL detection is over a few revolutions, i.e provides an average flux,
we have combined the two most  statistically significant XRT spectra and repeated the spectral analysis.
To account for the high z value,  all modeling has been carried out in the source rest frame.
A simple power law fit to this stacked spectrum still provides a flat photon index ($\Gamma$=1.31$\pm$0.09 ) and an acceptable 
$\chi^{2}$(46.9/53). However, inspection of the residuals to this model reveals  some  curvature possibly due to intrinsic 
absorption in the QSO rest frame.  Addition of this extra component  provides a significant fit improvement (99.98 \% confidence level using the F test), a  steeper spectrum
($\Gamma$= 1.53$\pm$0.16) and a column  density N$_{H}$=3$\pm$2 $\times$ 10$^{22}$ cm$^{-2}$.
Intrinsic absorption is a common property of radio loud/high redshift QSO (Page et al. 2005) and 
it is generally ascribed to the presence of absorbing material in the jet.
However, this interpretation poses some difficulties as jets are expected to be efficient in removing gas in their vicinity. An alternative  possibility is that the continuum is intrinsically curved (Tavecchio et al. 2007) and well described by a broken power law: indeed
this model provides an equally significant improvement in the fit, a break energy at 1.55$\pm$ 0.15 keV and a spectral flattening
below 1-2 keV of $\Delta$$\Gamma$= 0.5 ($\Gamma_{1}$=0.7$\pm$0.3 and $\Gamma_{2}$=1.2$\pm$0.4).  
Finally, a joint spectral fit to the Swift/XRT and  INTEGRAL/IBIS data was also attempted.
While we find a perfect match in spectral shape with $\Gamma$=1.5$\pm$0.1 (for an absorbed power law), the XRT data fall short of the INTEGRAL detection by a factor in the range 3-7 (much higher than the XRT/IBIS cross calibration constant which is typically around 1, Masetti et al. 2007) , implying some variability 
in the source flux between observations (see figure 3). Indeed, when the shorter XRT observation is also used ( see Table 1), a decrease in the source intensity  becomes evident over the 6 day period of the Swift observational campaign. Variability in QSO is not unexpected and can be used to further characterized this new gamma-ray source.
Assuming that the IBIS observation  represents the average state of the  source, we obtain rest frame 
luminosities$\footnote{We adopt H$_{o}$=71 km s$^{-1}$ Mpc$^{-1}$, $\Omega_{\Lambda}$=0.73 and $\Omega_{M}$=0.27}$
of 0.3 $\times$ 10$^{48}$ erg  s$^{-1}$ in the X-ray (2-10 keV) 
band,  2 $\times$ 10$^{48}$ erg  s$^{-1}$ at  hard X-rays  (20-100 keV) and 
5 $\times$ 10$^{48}$ erg s$^{-1}$ in the soft gamma-ray  (100-500 KeV)
interval, i.e. MG3 J225155+2217 is an X/gamma-ray lighthouse shining from the edge of our Universe.

\begin{figure}
\rotatebox{-90}  {\epsscale{0.75} \plotone{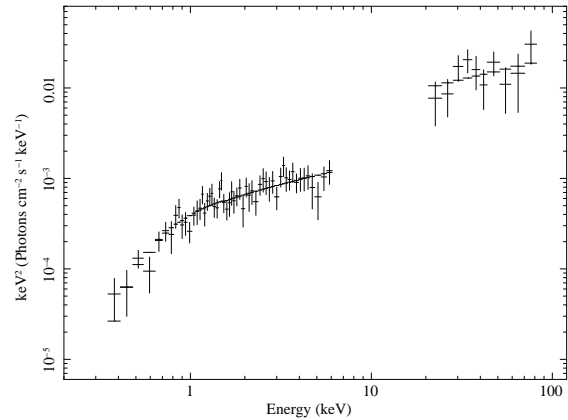}}
\caption  {Broad band spectrum of 
IGR J22517+2218/MG3 J225155+2217 fitted with an intrinsically absorbed power law (continuous line in figure): stacked XRT data
to the left and IBIS data to the right\label{fig3}} 
\end{figure}

\section{What type of QSO is IGR J22517+2218/MG3 J225155+2217?}
Although a search in the NED database only shows 10 references, the sparse information available 
are sufficient to broadly characterize this new INTEGRAL source.
The NVSS (NRAO VLA Sky Survey, Condon et al. 1998) image of MG3 J225155+2217 shows the source to be core dominated 
with no extended radio features and a 20 cm flux of 190 mJy. From the available data Kuhn (2004) estimated a radio loudness Log R=3.9 (well above the value used to define a source as radio loud) and a radio spectral index (F$_{\nu}$ $\propto$ $\nu^{\alpha}$)= -0.02 (flat enough to classify the object as a FSRQ). Incidentally, the chance probability of finding such a bright radio source in the INTEGRAL error box is 0.01, further supporting the association  between IGR J22517+2218 and the QSO.  
Optically,  MG3 J225155+2217  is a broad emission line AGN (FWHM $\ge$ 2000 km s$^{-1}$ 
in the source rest frames, Falco et al. 1998); it is  
further labeled as a narrow absorption line (NAL) system due to the presence of absorption lines 
that are significantly narrower than the emission lines. These absorption features are seen in around 10$\%$ of the 
spectra of high redshift QSO (Boksenberg 1978) and are either intrinsic to the source or  the result
of intervening gas between us and the QSO.  Observations indicate that radio loud objects with 
flat radio spectra tend to avoid strong NAL, intrinsic or otherwise, and this make MG3 J225155+2217 peculiar in this respect (Ganguly et al. 2001). In X/gamma-rays, the source is bright  with a hard X-ray spectral shape ($\Gamma$=1.5); a deficit of soft photons in the spectrum  can be either 
due to absorption local to the QSO or  to intrinsic spectral curvature.
The source is likely variable at high energies and possibly also at lower frequencies (Kuhn 2004).\\
Most of the above  properties strongly suggest that MG3 J225155+2217 is a bright blazar, in which the emission is 
relativistically beamed and the SED double peaked. 
In figure 4,  we construct the non-simultaneous rest frame SED of this enigmatic object 
by combining data from this work, from the HEASARC archive and from Kuhn 
(2004); to cover as many frequencies as possible we have also used upper 
limits obtained  from  the IRAS (Moshir et al. 1990) and EGRET (Hartmann et 
al. 1999) surveys. 
To deal with high energy variability, we have used in 
the SED the absorbed power law X/gamma-ray spectrum normalized to the IBIS 
flux. For comparison, we have also plotted in figure 4 the synchrotron inverse 
Compton model used to reproduced the SED of a  blazar at comparable distance
([HB89] 0836+710, z=2.17, Tavecchio et al. 2000) scaled to match 
both the high energy data and the redshift of MG3 J225155+2217.
The first striking feature of figure 4 is the extreme brightness of the 
source at X/gamma ray energies compared to the emission at lower 
frequencies: the  X-ray to radio flux ratio is $\sim$ 1000 (or $\alpha_{xr}$ 
$<$ 0.75), which makes it similar to high energy peaked blazars, i.e. those 
with the synchrotron peak in the X-ray band.  The second important point is 
the shape of the infrared to optical continuum which is at odds with the 
location of  the synchrotron peak  at infrared frequencies as generally 
observed in FSRQ like [HB89] 0836+710 .
Taken together, this indicates that this is a peculiar and rare object 
within the blazar population (Padovani 2007). Until recently, no high energy peaked 
FSRQ seemed to exist but some are now emerging such as the very  powerful 
z$\sim$4 FSRQ ROXA J081009.9+384757.0 with a possible synchrotron peak in the 
X-ray band (Giommi et al. 2007). If the peak we see at around 100 keV (or 
$\sim$ 400 keV in the source rest frame) is due to the synchrotron component 
then : a) the blazar sequence in its simple formulation (Fossati et al. 1998) 
would be severely questioned by this result; b) the observed  peak energy would be  the highest ever measured in a FSRQ meaning that no difference in this parameter exists between low and high luminosity blazars; c) the Compton peak would be 
well into the gamma-ray band making the source a good target for AGILE and  GLAST.
Obviously, the sparse and non-simultaneous data coverage still leaves open 
the possibility of a more classical double humped interpretation of the 
source SED; but even in this case the object is strange as not only the 
above peculiarities have to be explained, but in this case the Compton peak 
would be at MeV energies, i.e. at lower energies than typically observed in FSRQ. 
Either way, MG3 J225155+2217 is quite atypical for its class and an extreme object in the blazar population;
this makes it an interesting LABORATORY in where to test current blazars theories and so an object worth following up at all wavebands.

\begin{figure}
\plotone{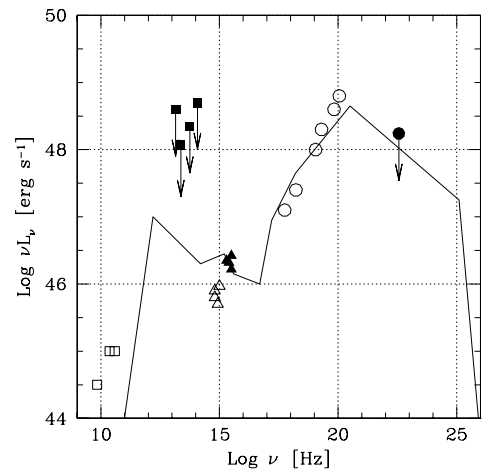} \caption  {Non simultaneous rest frame SED of IGR J22517+2218=MG3
J225155+2217. Data have been taken from 
HEASARC archive: radio (open squares),optical (full triangles); Moshir et al. (1990):far-infrared/IRAS (full squares);
Kuhn (2004): near-infrared (open triangles);this work: X-rays/XRT-IBIS (opens circle)
and Hartman et al. (1999):gamma-rays/EGRET (full circle)\label{fig4}. The line represents the SED of [HB89] 0836+710 normalized to MG3 J225155+2217 (see text) }
\end{figure}

\section{Conclusions}
Through X-ray follow up observations with Swift/XRT, we have been able to 
identify  the newly discovered INTEGRAL source, IGR J22517+2218 with MG3 
J225155+2217, a QSO at  z=3.668. This makes IGR J22517+2218
the most distant object so far detected by INTEGRAL and the second most 
distant blazar ever observed above 20 keV. The source has many peculiar features 
which deserve more in depth studies: it is  Aradio loud object with 
a flat spectrum; it hosts a NAL system in optical; 
it  shows  a spectral curvature in X-rays either intrinsic or due to high 
absorption in the source rest frame and it is  variable at high energies. 
The source SED is also unusual among  FSRQ as it is compatible with having 
the synchrotron peak in the X/gamma-ray band (i.e. much higher than 
generally observed) or alternatively with the Compton peak in the MeV range 
(i.e. lower than  typically measured).

\acknowledgments 
We acknowledge financial support from ASI (contracts I/R/008/07/0 and I/023/05/0) and 
PPARC (grant PP/C000714/1).
This research has made use of the NED NASA/IPAC Extragalactic Database (NED) 
operated by JPL (Caltech) laboratory and of the HEASARC archive provided by NASA's Goddard Space Flight Center.

\end{document}